\documentclass{mem}
\usepackage{natbib}\usepackage{txfonts}\usepackage{balance}
\usepackage{graphicx}
\usepackage[a4paper]{hyperref}
\idline{75}{282}
\begin{document}
\def\teff{$T\rm_{eff }$}
\def\kms{$\mathrm {km s}^{-1}$}

\title{
Diffuse UV Background:  GALEX Results
}

   \subtitle{}

\author{
Richard Conn \,Henry
          }

% \offprints{Richard Conn Henry}

\institute{
Henry A. Rowland Department of Physics \& Astronomy \\
The Johns Hopkins University\\
3400 North Charles Street\\
Baltimore, Maryland, 21218-2686 USA\\
\email{henry@jhu.edu}
}

\authorrunning{Henry }

\titlerunning{Diffuse UV Background}

\abstract{
A bright UV GALEX image in the direction of a dense high
galactic latitude interstellar dust cloud  is examined to test (and to reject) the idea that a bright \emph{extragalactic}
UV background radiation field exists.  A GALEX ``Deep Imaging Survey" image of a second
high latitude region (a region almost totally free of dust) shows a similar bright
 background, which, clearly, cannot be due to starlight scattered from interstellar dust.  I
speculate that the background is due to dark matter particles interacting with
 interstellar gas/dust nucleons.
}
\maketitle{}

\section{Introduction}

    \begin{figure}[htbp]  
       \centering
       \includegraphics[width=2.5in]{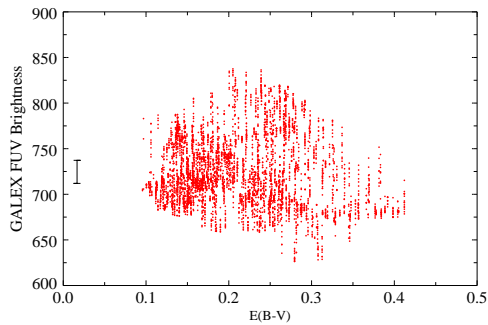}  
       \caption{SANDAGE: lack of correlation of FUV ($\sim1500$ \AA) GALEX brightness with E(B-V).}
       \label{fig:nocorr}
    \end{figure}

   \begin{figure}[htbp]   e
       \centering
       \includegraphics[width=2.5in]{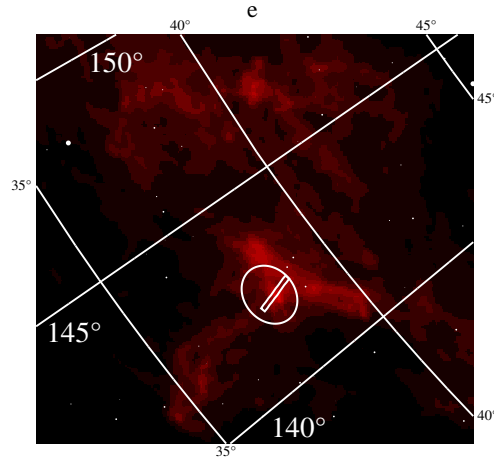}  
       \caption{  SANDAGE:  circle:  GALEX fov;  rectangle: Voyager spectrometer slit; 
       white dots:  TD1 stars; red color: IRAS E(B-V). $\delta=70\,\fdg4$
      }
       \label{fig:sky}
    \end{figure}
 
Sujatha, Murthy,  Karnataki, Henry, \& Bianchi (2009) have used 
the GALEX ultraviolet imagers  to study
the diffuse UV background at the high-galactic-latitude 
location ``SANDAGE," discovered by Allan Sandage (1976) to be extremely dusty (abscissa, Fig.~1).

    \begin{figure*}   %    Two column figure
   \centering
 \includegraphics[width=6in]{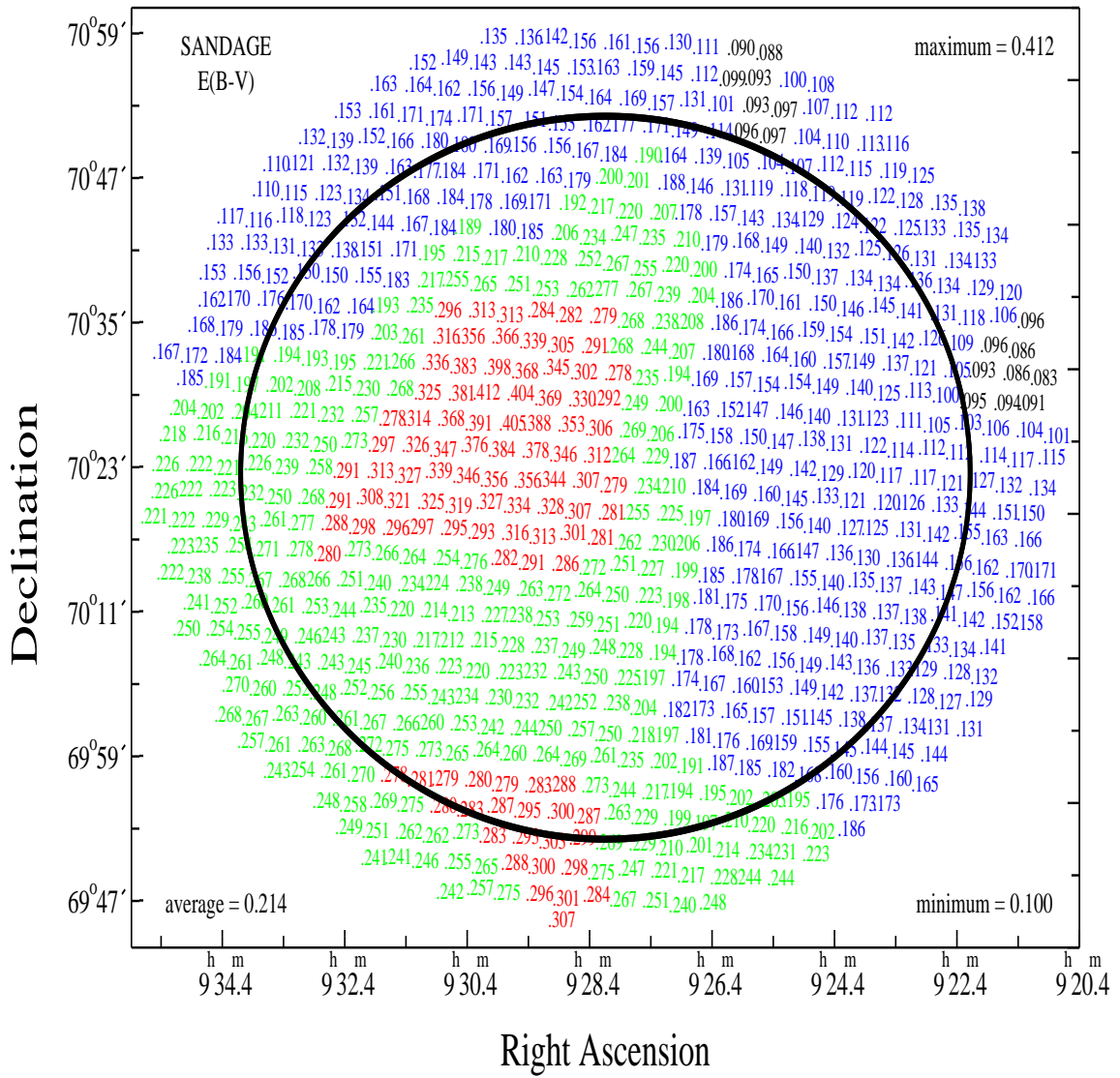}
   \caption{Values of E(B-V) for GALEX SANDAGE (Schlegel, Finkbeiner, \&  Davis 1998).
   The heavy line (one degree diameter) shows the portion of the data that can be interpreted.}
              \label{numbers}%
    \end{figure*}

They
showed that the large
observed signal ($\sim$750~photons cm$^{-2}$ s$^{-1}$ sr$^{-1}$ \AA$^{-1}$) might be due, in the main, to starlight scattered from dust.

  \begin{figure}[htbp]   
       \centering
       \includegraphics[width=2.5in]{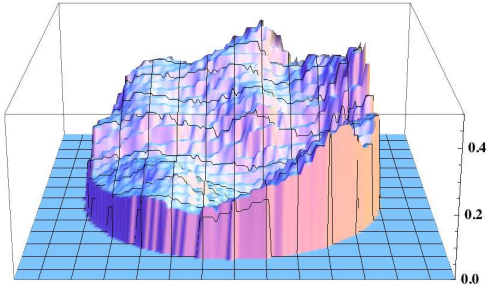}  
       \caption{Predicted GALEX FUV image if the background radiation comes from beyond the dust.
       (Any additional spatially uniform contributor would reduce the predicted contrast.)}
       \label{fig:modelSandage}
    \end{figure}

  \begin{figure}[htbp]   
       \centering
       \includegraphics[width=2.5in]{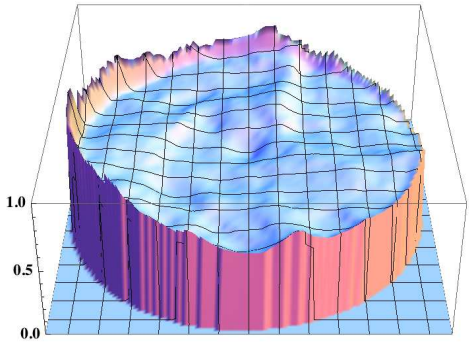}  
       \caption{The actual GALEX FUV image bears no resemblance to the prediction of Fig. 4.}
       \label{fig:sandageFUV}
    \end{figure}

In the present paper I use these same data to test the suggestion by Henry (1991, 1999) that the ubiquitous UV background
observed longward of ~1216 \AA\ is of extragalactic origin.  (This target, at  $b=38\,\fdg6$, was proposed to NASA expressly for the purpose of testing Henry's idea.)

  Fig.~3 gives the values of E(B-V) across SANDAGE. If I assume that the only signal from this direction is
  a uniform \emph{extragalactic} (i.e., from \emph{beyond} the $\sim100$ pc distance of the dust) background (and that the dust is forward-scattering) I predict 
  that GALEX should see the FUV image shown in Fig. 4.  In fact, the actual GALEX FUV image appears in Fig.~5.  
  It bears no resemblance to our prediction. Because the  dust
  is optically thick, starlight from foreground stars back-scattering from the dust could not ``fill the hole" that is predicted under the extragalactic hypothesis, which we can thus now
  rule out.

\section{A high latitude dust-free target}

 \begin{figure}[htbp]   
       \centering
       \includegraphics[width=2.5in]{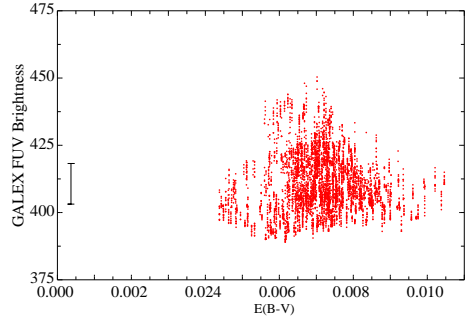}  
       \caption{A strong FUV background is seen at this $b=-79\,\fdg9$ ELAISS1$\_\,00$
       location (which is almost free of dust). Error bar, as in Fig. 1,
       shows statistical uncertainty in a typical data point. }
       \label{fig:numbers}
    \end{figure}

 \begin{figure}[htbp]   
       \centering
       \includegraphics[width=2.5in]{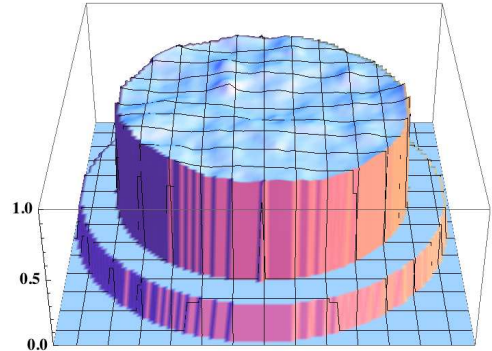}  
       \caption{The pedestal shows where data have been omitted from the ELAISS1$\_\,00$ image.}
       \label{fig:sandageFUV}
    \end{figure}

The GALEX FUV image of the diffuse background at SANDAGE is remarkable in appearance, but is nevertheless similar in appearance to 
\emph{every other} such image that I have seen.  
\emph{What can be the source of this strange background radiation?}  

Could it be starlight scattered from dust?
To find out, consider Deep Imaging Survey target ELAISS1$\_\,00$ which (Fig. 6) is almost \emph{free of} interstellar dust.  The
image (Fig. 7), despite the lack of dust, closely resembles that of SANDAGE.  I have written a simple fortran program to model
the expected dust-scattered starlight.  For a dust albedo of 0.28 and a Henyey-Greenstein scattering parameter $g= 0.61$ 
(Sujatha et. al. 2007), I calculate an expected background at 1500 \AA\  for this target of only 17~photons cm$^{-2}$ s$^{-1}$ sr$^{-1}$ \AA$^{-1}$.

 \begin{figure}[htbp]   
       \centering
       \includegraphics[width=2.5in]{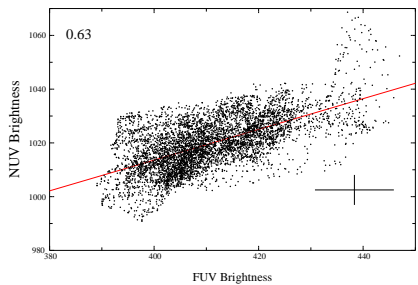}  
       \caption{There is significant correlation between the FUV (1350-1750 \AA)  and NUV (1750-2750 \AA) intensities for the ELAISS1$\_\,00$ images.
    }
       \label{fig:sandageFUV}
    \end{figure}

There  \emph{are} GALEX images in which dust-scattered starlight is clearly identifiable (Henry 2006), but the 
ELAISS1$\_\,00$ data prove that the \emph{ubiquitous} strong cosmic FUV background
is  \emph{not} due to dust-scattered starlight.  (And, as we have just seen, that background \emph{cannot} be extragalactic in its origin).

Fig. 8 shows the correlation between the NUV and FUV intensities for ELAISS1$\_\,00$.  
The correlation coefficient is 0.63. I have made no correction for the large
zodiacal light contribution to the NUV intensity (the FUV image has \emph{no} zodiacal light contamination).

\section{Source of the UV background}

 \begin{figure}[htbp]   
       \centering
       \includegraphics[width=2.5in]{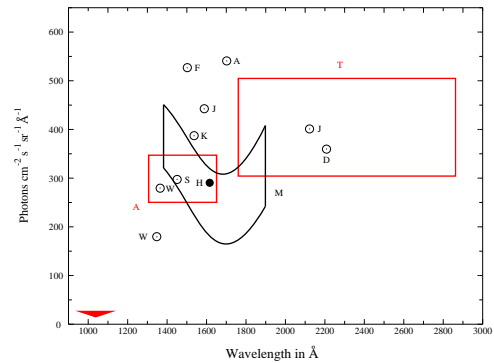}  
       \caption{
      Henry \& Murthy (1995) exhibited this strange spectrum of the diffuse UV background
      radiation.  The red triangle is the Voyager upper limit of
      30~photons cm$^{-2}$ s$^{-1}$ sr$^{-1}$ \AA$^{-1}$ that has been defended by Murthy et al. (2001).  The Aries rocket spectrum (A) by one of my students (Anderson et al. 1979) shows \emph{continuum} radiation, as does (at longer wavelengths) the Aries rocket spectrum (T) by another of
      my students (Tennyson et al. 1988); we corrected the Tennyson et al. observation meticulously  for airglow, and for zodiacal light.
       }
       \label{fig:numbers}
    \end{figure}

I began my investigation of the UV background in Henry (1973).  By 1995 the situation was as shown in Fig. 9---which
allows one to understand,  I hope, why
I believed the radiation longward of 1216 \AA\ to be redshifted recombination radiation from the intergalactic medium.  But, the SANDAGE data show that
idea to be wrong.  What, then, \emph{is} the source of the radiation?  It cannot be dark matter annihilation radiation, for the photon energy would be much
too high---also, most such radiation would originate \emph{beyond} the SANDAGE cloud.  I suggest that the radiation results from collisions of dark
matter particles with interstellar medium nucleons.  The weak interaction conveys energy to the electrically-charged quarks, which radiate.  See
the lattice gauge models of  Derek B. Leinweber:  http://www.physics.adelaide.edu.au/theory/$\rightarrow$
staff/leinweber/VisualQCD/Nobel/
Because the interaction cross section is expected to go as the square of the atomic number, hydrogen, helium, and metals would contribute about equally
to the background radiation.

 \begin{figure}[htbp]   
       \centering
       \includegraphics[width=2.5in]{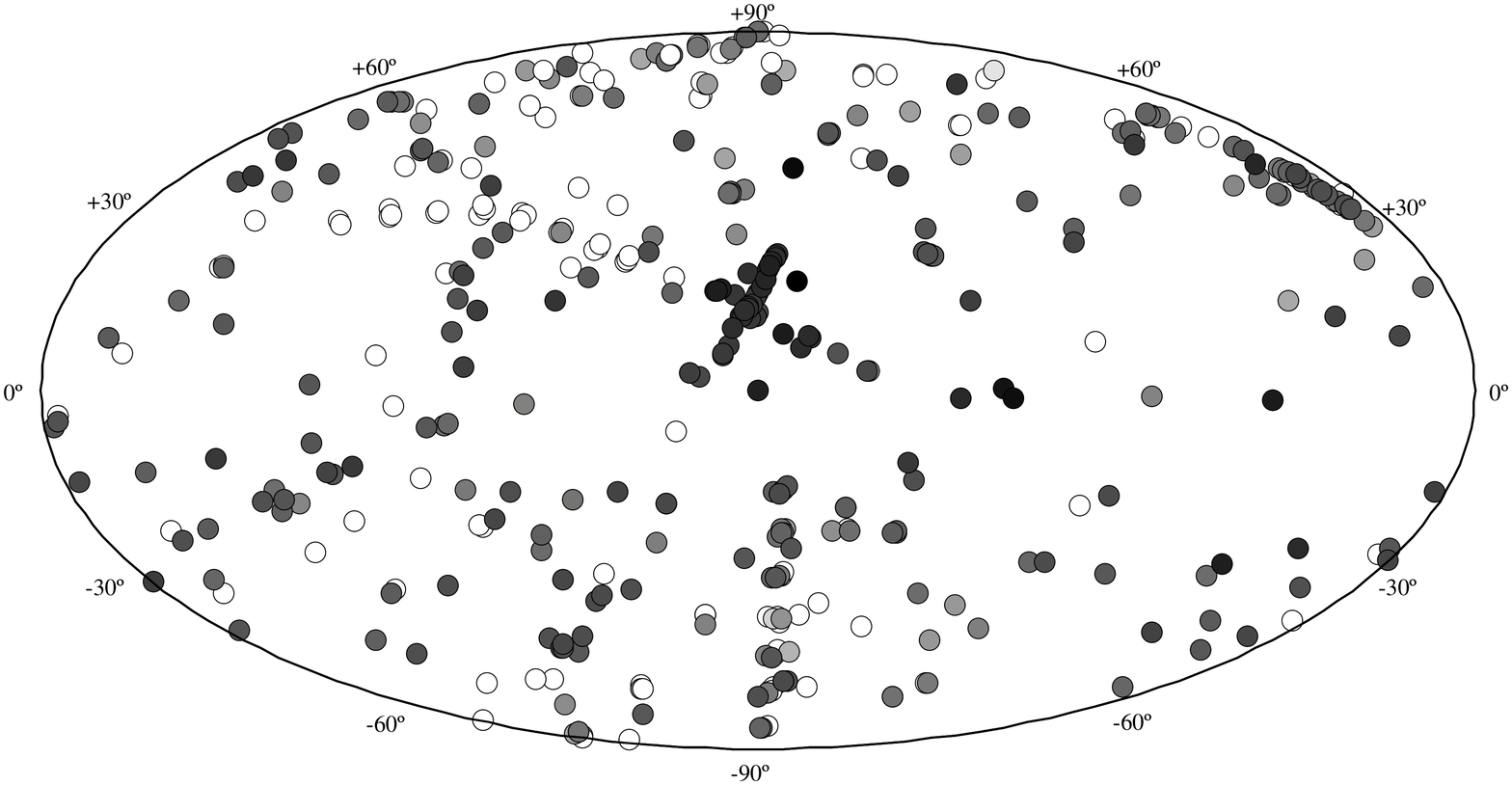}  
       \caption{
      This Galactic-coordinate map shows the Voyager (Murthy et al. 1999) diffuse UV background at $\sim$1000 \AA. Filled circles
      show dust-scattered starlight; open circles have only an upper limit (as in Fig. 9).  So, at $\sim$1000 \AA\ the UV background shows \emph{high contrast},
      while the GALEX images at $\sim$1500 \AA\ show, thus far, \emph{no} locations with low background.
       }
       \label{fig:numbers}
    \end{figure}

   \begin{figure*}   %    Two column figure
   \centering
 \includegraphics[width=6in]{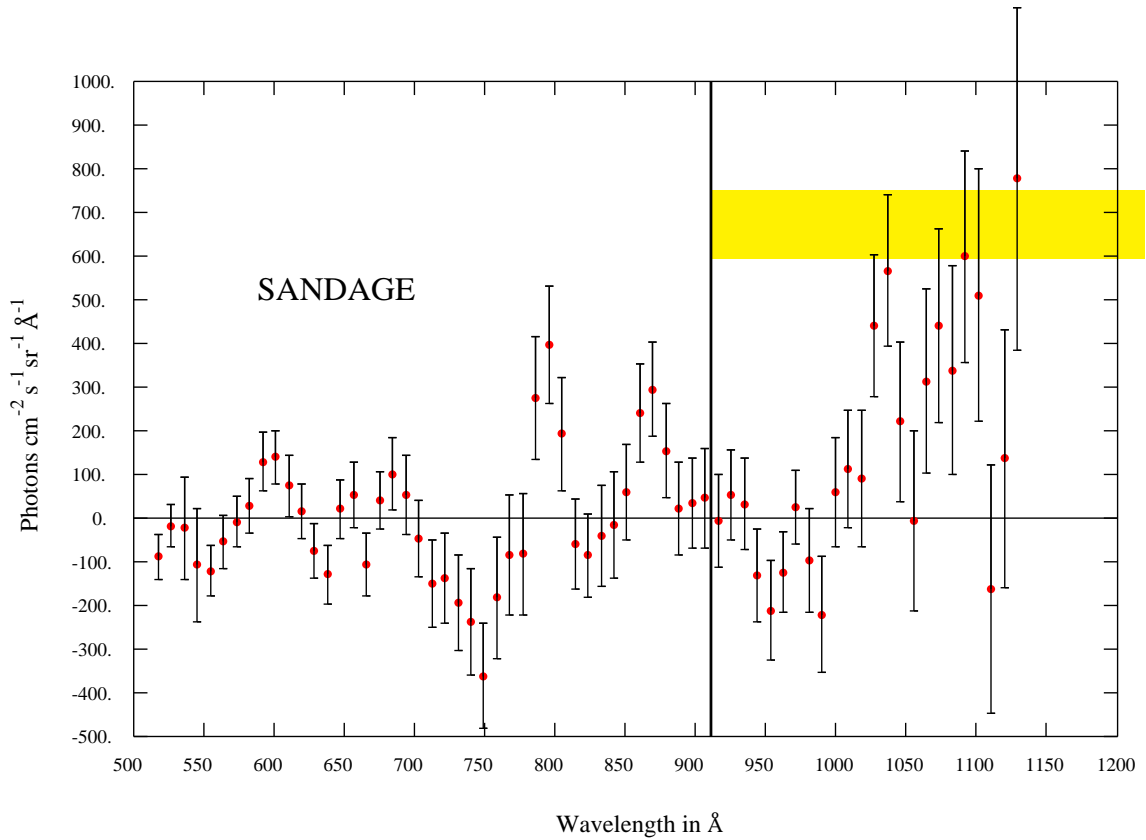}
   \caption{
   Voyager UV spectrum of the diffuse cosmic background in the direction of SANDAGE (Murthy et al. 1999).  Murthy et al. assumed that there is no 
   cosmic signal short of the interstellar hydrogen ionization edge.  We see that for SANDAGE there is \emph{also} no cosmic signal \emph{longward} of that edge, but 
   beyond $\sim$1000~\AA\,
   a signal appears which rises to meet the observed GALEX SANDAGE image intensity (yellow band) of Fig.~1.  This strongly suggests that there is 
   minimal contamination in the GALEX image.
   }
              \label{Voyager}
    \end{figure*}

\section{Contamination in GALEX images}

Signal in both the FUV and NUV GALEX imagers (while pointed at a fixed target) always decline until local midnight, and then rise as dawn approaches (see Fig. 2
of Sujatha et al. 2009).  Murthy (private communication) has discovered that the magnitude of the decline correlates with solar activity.  The GALEX images
in Fig.~5 and Fig.~7 \emph{look like} they could be 100\% contamination.  (The NUV images are seriously contaminated with zodiacal light, so in the present paper I
have largely confined my attention to the FUV images.)  The FUV imager may transmit some OI 1356 \AA\ airglow, and there could also be a background due
to particle fluorescence in the window.  Each of these would be expected to vary with both time and solar activity.  

In the next section, however, I will demonstrate that only a \emph{small fraction} of
what appears in Fig.~5 and Fig.~7 can, in fact, be due to contamination:  we are mostly seeing cosmic signal.

\section{Discussion}

In Fig. 11, I plot the Voyager spectrum for SANDAGE (Murthy et al. 1999).  Fig. 2 locates the Voyager spectrometer slit
(28 \AA\ resolution) on SANDAGE.  The Voyager spacecraft was powered by an RTG nuclear source that
contributed a strong wavelength-independent background to the UV spectrometers.  Murthy et al. \emph{subtracted}
a wavelength-independent background--- just enough to 
give zero signal shortward of 912~\AA.  But now notice that from 912~\AA, to beyond 1000~\AA,  there is \emph{also} no signal:  a clear
indication that \emph{a)} the subtraction of background has been done right, and \emph{b)} there truly \emph{is} no celestial background
radiation just longward of the interstellar hydrogen ionization limit.  Finally, notice that the Voyager spectrum then \emph{rises} toward the level (yellow 
band) that we see in our GALEX image of this target.

Our GALEX SANDAGE image covers the range 1350~\AA\ to 1750~\AA\ with average signal about 750 photon units (top of yellow band) which
should be reduced to $\sim$600 photon units (bottom of yellow band) to allow for contamination, following Murthy's correlation.  
This Voyager spectrum surely rules out any larger contribution
 from geophysical contamination to the GALEX image.

The ELAISS1$\_\,00$ data shown in Fig. 6 were acquired in 2003, when solar activity was high.  A similar plot for observation
of the same target in late 2006, when solar activity was much lower, shows a decline to about 400 photon units---hardly a substantial decrease.

Finally, the data shown in Fig. 9 include \emph{no} GALEX data, and are from a wide variety of instruments.  Like GALEX, they show a background
of order 400 photon units in every case.  

So, there is a ubiquitous cosmic background longward of about 1000~\AA, and it has the appearance that is 
exhibited in Fig. 5 and Fig. 7---which look unlike any other astronomical photographs.

\section{Conclusions}

I think in Figs. 5 and 7 we are, for the first time, seeing the bulk of the interstellar medium by means of radiation (``radiative corrections") from dark matter 
particles impacting interstellar medium nucleons.  Assuming that the nucleons are destroyed in the interaction (which may not be 
the case), some thousands of UV photons must be emitted per interaction, if the interstellar medium is not to be depleted over the age 
of the universe.

Our FUV background, we know, continues into the NUV (Tennyson et al. 1988) and, perhaps, even into
the visible (Henry 1999)---most of the photons in the Hubble Deep Field image come, not from the galaxies, but from a diffuse optical background, which, I suggest,
is the continuation to the optical of the spectrum that appears in Figs.~9 and 11.

\begin{acknowledgements}
I am grateful to my colleagues, and for support from the United States National Aeronautics and Space Administration 
(NASA).
\end{acknowledgements}

\bibliographystyle{aa}

\end{document}